# Annotation of epidemiological information in animal disease-related news articles: guidelines

Sarah Valentin[1,2,3], Elena Arsevska[1,3], Aline Vilain[4],
Valérie De Waele[5], Renaud Lancelot[1,3], Mathieu Roche[2,3]

[1] UMR ASTRE (Unit for Animals, Health, Territories, Risks and Ecosystems), University of Montpellier, CIRAD, INRAE, Montpellier, France

[2] UMR TETIS (Land, Environment, Remote Sensing and Spatial Information), University of Montpellier, AgroParisTech, CIRAD, CNRS, INRAE, Montpellier, France

[3] French Agricultural Research for Development (CIRAD), France

[4] Veterinary Epidemiology Service, Departement of Epidemiology and Public Health, Sciensano, Brussels, Belgium

[5] Department of Environmental and Agricultural Studies, Public Service of Wallonia, B5030 Gembloux, Belgium

## 1. Introduction

The event-based surveillance (EBS) gathers information from a variety of data sources, including online news articles and social media. Contrary to the data from the formal reporting, the EBS data is not structured, and its interpretation can overwhelm EI capacities in terms of available human resources. Therefore, there has been a development of diverse EBS systems that automatically process (all or part) the acquired non-structured data. These EBS systems GPHIN (Blench 2008), HealthMap (Freifeld *et al.* 2008), MedISys (Alomar *et al.* 2016), ProMED (Carrion and Madoff 2017) and PADI-Web (Arsevska *et al.* 2018) can use annoted data in order improve surveillance systems (Valentin *et al.* 2020).

This paper describes a method for annotation of epidemiological information in animal disease-related news articles. The annotation guidelines are generic and aim to embrace all animal or zoonotic infectious diseases, regardless of the pathogen involved or its way of transmission (e.g. vector-borne, airborne, by contact). The framework relies on the successive annotation of all the sentences from a news article. The annotator evaluates the sentences in a specific epidemiological context, corresponding to the publication of the news article.

## 2. Annotation process

The annotation process follows the above steps:
1. Reading of the news article metadata (title, publication date, and source),
2. Reading of the whole news article as many times as needed to understand the context fully and to identify the epidemiological information that it contains,
3. Annotation of the *Event type* label for each sentence,
4. Annotation of the *Information type* label for each sentence, except the ones labelled as "**Irrelevant**" at the previous step.

The following provides important points that need to be considered during the annotation:

- **Double level annotation**

This annotation framework relies on the attribution of two labels per sentence: *Event type* and *Information type*. This double-level approach aims to assign consistent information type labels, whatever the temporality or the polarity of the sentence (recorded in *Event type* label). All *Information type* labels can be encountered with all *Event type* labels, apart from the *Information type* label "**General epidemiology**", specific to the *Event type* label "**General**" (Section 3.2).

- **Date of reference**

When determining the *Event type* label, the annotator uses the publication date of the news article as it was the current date.

- **Unit of annotation**

This annotation framework is sentence-based: the annotator labels each sentence from the news article separately. However, the meaning of a sentence depends on the entire article content. For instance, considering the two following sentences:

> *"Since the outbreaks, the ministry has taken many disease control and prevention measures."* [1]
>
> *"It has cooperated with public security departments to trace its origin."* [2]

Sentence [2] cannot be fully understood by itself. In this annotation frame, the annotator uses the whole context (here, the previous sentence), to read sentence [2] as "*[the ministry] has cooperated with public security departments to trace [the disease 's] origin.*"

- **Multi-topic sentences**

In this framework, the annotator has to choose one single label per level and per sentence. As some sentences may contain information belonging to several *Information type* categories, the annotator must decide what the primary information (label) is. We provide some indications to help to overcome some typical cases of multi-topic sentences in Section 3.3.

- **Negation**

The negation is not taken into account in this framework, i.e. the category of a sentence is the same whether the sentence is affirmative or not.

## 3. Label description

### 3.1. Event type labels

*Event type* label indicates the relation between the sentence and the epidemiological context (at the publication date). This label aims to differentiate sentences referring to the current/new situation ("**Current event**" and "**Risk event**") from sentences referring to old outbreaks ("**Old event**") or general information ("**General**"). Sentences which do not contain any epidemiological information are considered as irrelevant ("**Irrelevant**").

*Event type*
- **Current event**: this class includes sentences related to the current situation. There are four major groups of sentences that are considered as "current":
  1. Sentences related to a current event (main event). For instance, "*South Korea confirms two new African swine fever cases.*"
  2. Sentences related to recent events (relatively to a main stated event). Those events generally occurred at a nearby location and/or within a short time window around the main event. For instance, "*On Saturday, similar infections were found in 30 pigs on a farm in the Huangpu district of Guangzhou.*"
  3. Sentences containing an aggregation of events between a prior date and a recent/current date. For instance, "*According to data from the Council of Agriculture, 94 poultry farms in Taiwan have been infected by avian flu so far this year.*" In the previous sentence, the temporal expression "*so far this year*" indicates a relationship between the start of the outbreak and the publication date.
  4. Sentences describing the recent/current epidemiological status of a disease within an area. For instance, "*In recent months, the disease has spread more rapidly and further west, affecting countries that were previously unscathed.*"
  5. Sentences about events that will definitively occur in the future. In general, this category includes the direct consequences of an event, such as control measures that will be taken. For instance, "*All pigs in the complex will be killed, and 3km and 10km protection and surveillance zones will be installed.*"

- **Old event:** This class includes sentences about events that provide a historical context for the main event. Those sentences contain explicit references to old dates, either absolute ("*In 2007*") or relative ("*Back in days*"). This category includes two groups of sentences:
  1. Sentences related to an old event. For instance, "*The most recent case of the disease in the UK came in 2007.*"
  2. Sentences containing an aggregation of events between two past dates. For instance, "*During last year, 132 cases were recorded across the country*".

3. Sentences describing the past epidemiological status of a disease within an area. For instance, *"Between 2006 and 2010, BTV serotype 8 reached parts of north-western Europe that had never experienced bluetongue outbreaks previously."*

- **Risk event:** This class includes all sentences referring to hypothetical events. These sentences are generally about an area at risk of introduction or diffusion of a pathogen. This category includes:
  1. Sentences about an unaffected area expressing concern or preparedness. For instance, *"Additional outbreaks of African swine fever are likely to occur in China, despite nationwide disease control and prevention efforts."*
  2. Sentences about an area where the disease status is unknown. For instance, **"***If the outbreak is verified, all pigs at the feeding station will have to be culled, Miratorg said."*

- **General:** This class includes general information about a disease or a pathogen. Classically, the sentences describe the disease's hosts, its clinical presentation and pathogenicity. For instance, *"Bluetongue is a viral disease of ruminants (e. g. cattle, sheep goats, deer)."*

- **Irrelevant:** This class includes sentences that don't contain any epidemiological information. This group includes, for instance, disease-unrelated general facts (*"Pig imports from Hungary only represented about 0. 64 per cent of all pork products to the UK in 2017."*) or article news artefacts (*"Comments will be moderated."*).

### 3.2. Information type

*Information type* label qualifies the type of epidemiological information. It aims to distinguish: (i) sentences detailing the circumstances of a specific event or the status of a disease in an area ("**Descriptive epidemiology**" and "**Distribution**"), (ii) preventive or control measures against a disease outbreak ("**Preventive and control measures**"), (iii) event's economic and/or political impacts ("**Economic and political consequences**"), (iv) event's suspected or confirmed route of transmission ("**Transmission pathway**"), (v) sentences expressing a concern and/or facts about risk factors ("**Concern and risk factors**"), and (vi) general information about the epidemiology of a pathogen or a disease ("**General epidemiology**").

- **Descriptive epidemiology:** This class includes sentences containing the standard epidemiological indicators (e.g. disease, location, hosts, dates) used to describe an event. It includes:
  1. The epidemiological description of the event. For instance, *"Cases of African swine fever (ASF) have been recorded in Odesa and Mykolaiv regions."*

2. Information about the pathogenic agent and its strain (if relevant). For instance, *"Results indicated that the birds were infected with a new variety of H5N1 influenza."*
3. The clinical signs associated with the disease or the suspected pathological event, as well as element regarding its morbidity and mortality. For instance, *"The remaining buck appears healthy at this time and is showing no clinical signs associated with the disease."*

- **Distribution:** This class contains the sentences giving indications on the status of a disease in a specific area (i.e. an administrative unit, a country, a continent). It includes:
    1. Sentences describing the epidemiological status. For instance, *"In recent months, the disease has spread more rapidly and further west, affecting countries that were previously unscathed."*
    2. Sentences containing an aggregation of events between a past date and a recent/current date. For instance, *"According to data from the Council of Agriculture, 94 poultry farms in Taiwan have been infected by avian flu so far this year."*

- **Preventive and control measures:** This class includes sentences describing:
    1. Preventive measures, i.e. all the sanitary and physical actions taken to avoid the introduction of a disease into an unaffected area. For instance, *"ASF: France about to end the fencing in the borderland with Belgium."*
    2. Control measures, i.e. all the sanitary and physical actions that are taken to control spread and eradicate a pathogen once introduced into an area (e.g. vaccination, slaughtering, disinfection, zoning, etc.). For instance, *"All the infected animals have been killed, and the area has been disinfected."*
    3. Instructions/recommendations, i.e. actions for both preventive and control measures. For instance, *"Hunters, travellers, and transporters are asked to take extra care concerning hygiene."*

- **Transmission pathway**: This class includes the sentences indicating the origin (source) of the pathogen or the route of transmission. For instance, *"The authorities suggest that the highly contagious virus might have been spread by a river"*.

- **Concern and risk factors:** This class includes sentences indicating a risk of introduction or spread of disease into an area. We include two types of sentences in this group:
    1. Sentences claiming or suspecting one or several risk factors, according to the outbreak definition (i.e. individual, behavioural and environmental characteristics associated with an increased occurrence of disease). For example, *"A recent wave of inspections revealed 4,000 different biosecurity violations on farms and Gosvetfitosluzhba warned that this could result in further outbreaks soon."*

2. Sentences semantically expressing fear or concern regarding:
    (i) The hypothetical introduction of a pathogen into an unaffected area. For instance, *"ASF is a real threat to the UK, she said."*
    (ii) The worrying evolution of an epidemiological situation. For instance, *"Several countries are affected, alarming governments and pig farmers due to the pace at which the disease has spread."*

- **Economic and political consequences:** This category includes all references to the direct or indirect financial or political impacts of an outbreak on a geographical area. It consists of the result of preventive and control measures. For instance, *"Gorod estimated that financial losses due to ASF could amount to €17 million to Latvia's industry in 2017."*

- **General epidemiology:** This category is only used for the sentences labelled as "**General**" as the *Event type* level. It merges the classes "**Event description**" and "**Transmission pathway**" described above. In this particular *Event type* level, those two categories include the description of a disease's hosts, pathogenicity and way of transmission. For instance, *"The virus is transmitted by midge bites, and it does not affect humans."*

### 3.3. Multi-topic sentences

As sentences can contain several types of information, therefore corresponding to distinct *Information type* labels, we provide two rules to help the annotator in its choice:
- If a category (label) is the consequence of another one, the annotator should select the first one. For instance, if a sentence describes both a control measure and its economic effects, the sentence should be labelled as "**Protective and control measures**".
- Some categories are more informative than other ones, i.e. we prioritise them against the others. Categories "**Concern and risk factors**" and **"Transmission pathway"** provide highly valuable information to assess the risk of emergence or spreading of disease. Therefore, the annotator should select those two labels primarily if identified into a multi-topic sentence.

Following these two rules, Table 1 shows the main labels for frequently encountered multi-topic cases.

**Table 1. Resolution of multi-topic sentences in typical cases.**
DE: Descriptive epidemiology, PCM: Protective and control measure, EPC: Economic and political consequences, TP: Transmission pathway

| Sentence topic | Example | Possible labels → main label | Rationale |
|---|---|---|---|
| Description of an event and its control measures. | *The Philippines confirms African swine fever, culls 7000 pigs.* | DE, PCM → DE | Control measures are consequences of the event. |
| Sanitary bans. | *Russia's agriculture authorities introduced temporary restrictions on pig and pork imports from Hungary due to an outbreak of the disease.* | PCM, EPC → PCM | Economic consequences consequences of the ban. |
| Description of an event and its source. | *The strain detected in China is similar to the one that infected pigs in eastern Russia last year, but there is no conclusive evidence of the outbreak's source, it said*. | DE, TP → TP | Transmission pathway category prevails over the other types. |

## Acknowledgements

We thank the members of the French Epidemic Intelligence Team in Animal Health for their constructive comments during the development of PADI-web. This work has been supported by the French General Directorate for Food (DGAL), the French Agricultural Research Centre for International Development (CIRAD), the SONGES Project (FEDER and Occitanie), and the French National Research Agency under the Investments for the Future Program, referred as ANR-16-CONV-0004 (#DigitAg). This work has also been funded by the "Monitoring outbreak events for disease surveillance in a data science context" (MOOD) project from the European Union's Horizon 2020 research and innovation program under grant agreement No. 874850 (https://mood-h2020.eu/).

The research that yielded this guide was also funded by the Belgian Federal Public Service Health, Food Chain Safety and Environment through the contractRT 18/2 MORISKIN 1, related to the research project Moriskin (Method for threat analysis in the context Of the RISK of emergence or re-emergence of INfectious animal diseases), coordinated by Sciensano (Institute of Public and Animal Health).

# References

Alomar, O., Batlle, A., Brunetti, J. M., García, R., Gil, R., Granollers, T., et al. (2016). Development and testing of the media monitoring tool MedISys for the monitoring, early identification and reporting of existing and emerging plant health threats. *EFSA Supporting Publications*, *13*(12). https://doi.org/10.2903/sp.efsa.2016.EN-1118

Arsevska, E., Valentin, S., Rabatel, J., de Goër de Hervé, J., Falala, S., Lancelot, R., & Roche, M. (2018). Web monitoring of emerging animal infectious diseases integrated in the French Animal Health Epidemic Intelligence System. *PLOS ONE*, *13*(8), e0199960. https://doi.org/10.1371/journal.pone.0199960

Blench, M. (2008). Global public health intelligence network (GPHIN). In *8th Conference of the Association for Machine Translation in the Americas* (pp. 8–12). https://pdfs.semanticscholar.org/7d88/e623aa6ca78510e0093e17e2e00db39bdad5.pdf. Accessed 13 October 2017

Carrion, M., & Madoff, L. C. (2017). ProMED-mail: 22 years of digital surveillance of emerging infectious diseases. *International Health*, *9*(3), 177–183. https://doi.org/10.1093/inthealth/ihx014

Valentin, S., Arsevska, E., Mercier, A., Falala, S., Rabatel, J., Lancelot, R., Roche, M. (2020) PADI-web: An Event-Based Surveillance System for Detecting, Classifying and Processing Online News. *Human Language Technology. Challenges for Computer Science and Linguistics - 8th Language and Technology Conference, LTC 2017, Revised Selected Papers. Lecture Notes in Computer Science 12598, Springer,* 87-101. https://doi.org/10.1007/978-3-030-66527-2_7